\documentclass[12pt]{iopart}

\usepackage{iopams}  
\begin{document}

\title[An introduction to generalised functions in periodic quantum theory]{An introduction to generalised functions in periodic quantum theory}

\author{Ian G Fuss$^1$ and Alexei Filinkov$^2$}

\address{$^1$ School of Electrical and Electronic Engineering, University of Adelaide, Adelaide SA 5005, Australia}
\address{$^2$  School of Mathematical Sciences, University of Adelaide, Adelaide SA 5005, Australia}
\ead{\mailto{ifuss@eleceng.adelaide.edu.au}, \mailto{alexei.filinkov@adelaide.edu.au}}
\begin{abstract}
\noindent
A proof that minimum uncertainty states of the simplest periodic quantum system exist in a state space that is represented by a Colombeau algebra of generalised functions but not in Hilbert space or in the space of Schwartz distributions is given. There are two significant generalisations of Hilbert space that lead to the special Colombeau algebra. The first step is to a vector space of Schwartz distributions that contains the eigenstates of operators with continuous spectra, the second is to a Colombeau algebra that provides a solution for a differential equation with non-constant coefficients. 

From the perspective of Colombeau algebra a rigged Hilbert space of Schwartz distributions can be understood as a representation of the linear component of the algebra. The utility of this representation is illustrated by a proof that generalised eigen-decompositions exist for linear operators with  periodic continuous spectra. 
\end{abstract}

\vspace{2pc}
\maketitle

\section{Introduction}
Quantum mechanical theories of periodic systems provide a foundation for our understanding of physical phenomena such as elementary particle fields, collective modes in condensed matter and molecular vibration. The purpose of this paper is to explore the question: what is an appropriate topological structure for a representation of the state space of  a periodic quantum theory. We demonstrate that rigour necessitates a  quantum theory with a state space of generalised functions for the simplest periodic quantum system. More than this we show that working with a periodic quantum theory based on Colombeau generalised functions, that can be multiplied and differentiated, is as simple and elegant as working with a quantum theory that is based upon infinitely differentiable classical functions \cite{Grosser}. Although our considerations eschew complex structures such as those required for vector gauge fields \cite{Weinberg1}, \cite{Weinberg2} and mathematical subtleties such as arise with non-selfadjoint operators \cite{Hall} our intention in providing a study that is focussed on the significance of periodicity for the mathematical representation of quantum mechanics is ultimately to contribute to a deeper understanding of these more complex and subtle quantum theories and aid in their development. 

The choice of periodic systems as a means of introducing generalised function quantum theories has a number of mathematical and pedagogical advantages. A key advantage stems from the fact that given a pair of complementary kinematic observables one of the observables is represented by an operator with a discrete spectrum and thus has eigenstates in a Hilbert space. In  this paper we seek to compound this advantage by using the simplest periodic quantum system the planar rotator \cite{Schrodinger} as an exemplar. For the planar rotator the angular momentum observable is represented by an operator with discrete eigenvalues and their corresponding eigenfunctions are complex exponentials. These eigenfunctions point to the next advantage of periodic systems and that is that all the states of a periodic quantum system whether these are represented by classical or generalised functions can be expressed as a Fourier series or parametrised families of them. We exploit the properties of these Fourier series throughout this paper in order to transform what can be abstract ideas associated with generalised functions into simple properties that are associated with Fourier series.

 The logical path that is followed in this paper of developing quantum theoretical descriptions with larger state space topologies traces a trajectory that closely parallels the historical processes associated with quantum theory and the theory of generalised functions. Early in the development of quantum mechanics it was realised by von Neumann \cite{von Neumann} and Dirac \cite{Dirac} that Hilbert space was unable to provide a simple and elegant expression of quantum mechanics due to its inability to provide a simple representation of the eigenstates that correspond to points within the continuous spectrum of self adjoint operators. Dirac addressed this issue operationally by introducing his heuristically based delta function at the end of the 1920s \cite{Dirac1} and his bra ket notation in the late 30s \cite{Dirac2}. In parallel the foundations of the mathematical theory of generalised functions were laid by Sobolev in his 1936 study of  the hyperbolic solutions for the Cauchy problem \cite{Sobolev}. In the more than 80 years since this initial work numerous contributions have brought us closer to a mathematically rigorous and physically adaptive description of quantum theory. A precise mathematical meaning was given to the Dirac delta function and other generalised functions as continuous linear functionals over a space of test functions by Schwartz \cite{Schwartz}. These ideas were generalised by Gelfand \cite{Gelfand}, to include self adjoint operators, via the concept of rigged Hilbert space that connects the analytic properties of a dense subspace of Hilbert space with its topological dual containing classical functions and continuous linear functionals. However Schwartz distributions \cite{SchwartzI} are not able to be multiplied freely a fact that is "inconsistent with many basic equations of [quantum] physics" \cite{Colombeau}. In order to solve these equations, that include linear differential equations with nonconstant coefficients, we need to preserve nonlinear information within their generalised function solutions. The associative and commutative differential algebras of generalised functions that were developed by Colombeau in the 1980s permit the maintenance of this information  \cite{Grosser} \cite{Colombeau1}, \cite{Colombeau2}. 

\section{Structure}
This paper is structured around some of the mathematical imperatives that arise from an analysis of the planar rotator. This model system is introduced in section 3.  In section 4 the kinematic operators for the planar rotator that correspond to the angle and angular momentum observables are examined with respect to the Hilbert space of square integrable functions on the unit circle. The angular momentum observable operator has a discrete spectrum with eigenfunctions in this Hilbert space and has a spectral decomposition on it with a form that we consider paradigmatic for all operators of periodic quantum systems. However the Hilbert space does not provide simple representations of the eigenfunctions of the angle operator a fact that is linked to its continuous spectra. 

A search for a state space that contains simple representations of the eigenfunctions of the angle operator is pursued in section 5 of the paper by examining the rigged Hilbert space of linear functionals that are defined on the space of infinitely differentiable functions on the unit circle. This rigged Hilbert space is shown to provide a locale for the eigenfunctions of the angle operator and eigenfunctions for differing eigenvalues are shown to be orthogonal in a distributional sense in an appendix. A constructive proof that generalised spectral decompositions of continuous periodic spectral operators exist for the rigged Hilbert space on the unit circle is given, thus fully implementing one of Dirac's heuristic ideas that is associated with his bra ket formalism. These spectral decompositions are examples of the so called eigenoperators that were first introduced into quantum theory by Roberts \cite{Roberts}, \cite{Roberts1} and they mirror the form of the spectral decompositions of operators with discrete spectra in Hilbert space. Although the rigged Hilbert space provides a good account of each of the kinematic operators and Borel functions of them it fails to provide the analytic tools needed to explore the consequences of their nonabelian product.

In section 6 of the paper we examine one of the fundamental implications of the fact that the angle and angular momentum operators do not commute. This implication is that we are unable to make precise simultaneous measurements of the angle and angular momentum of the planar rotator. This leads us to ask whether minimum uncertainty functions exist for the planar rotator. It has been shown that these states do not exist in the Hilbert space of square integrable functions on the unit circle \cite{Holevo} and we show that they do not exist in the rigged Hilbert space either. The failure of the rigged Hilbert space to provide a locale for the minimum uncertainty states of the planar rotator is a result of its linearity. A further topological extension of the planar rotator's state space to include the nonlinear functions of the special Columbeau algebra on the unit circle enables a proof that minimum uncertainty states do exist in this state space. We note that an association can be made between these minimum uncertainty states and the Dirac measure on the unit circle and that this association further clarifies the need to preserve nonlinear information in a generalised function in order to provide a solution to an equation with nonconstant coefficients.

\section{The planar rotator}
We consider the simplest periodic system the planar rotator as an  exemplar of periodic systems. The planar rotator can be modelled by a mass $m$ that moves on a circle of radius $R$. The potential energy of the mass is constant and may be taken to be zero. Hence the Hamiltonian for this fixed axis rotating body can be written as 
\[
H =  \frac{1}{2 I} J^2
\]
where $I = mR^2$ is the moment of inertia of the rotator and $J$ is its angular momentum \cite{Goldstein}. Thus the time independent Schrodinger equation for the planar rotator can be written in polar coordinates as  
\begin{equation}\label{eq1}
    H\psi (\theta) =  \frac{1}{2 I} \bigg( -i\hbar \frac{d}{d \theta}\bigg)^2 \psi (\theta) = E\, \psi (\theta)
 \end{equation}
 where $\theta$ is the angle coordinate, $E$ is the energy associated with the state $\psi$ and quantisation has been invoked by promoting the angular momentum to an operator and assigning to it the rule $J = -i\hbar \frac{d}{d \theta}$ with Planck's constant $\hbar$. This differential equation has the exponential solutions
\[
    \psi_k(\theta)  = A e^{i k \theta}\,, \ \theta\in [-\pi,\pi]\,,
\]
where $k = \pm \sqrt{\frac{2 \mathcal{I} E}{\hbar^2}}$ and $A$ is an arbitrary constant. A basic postulate of quantum mechanics is that the wave function of a particle without spin is single valued \cite{Debnath},  thus the quantum eigenfunctions of equation (\ref{eq1}) obey the condition $\psi_k(\theta + n\, 2\pi) = \psi_k(\theta)$ for $n\in\mathbb{Z}$ that implies $k\in\mathbb{N}$. Hence the elements of the set 
\[
    \bigg\{ \psi_k = e_k = \frac{1}{\sqrt{2\pi}}e^{i k \theta}\bigg\}_{k=-\infty}^\infty \, .
\]
are eigenfunctions of the planar rotator.

 \section{Kinematic operators on Hilbert space}
The complementary pair of kinematic operators of the planar rotator the angle operator  $\Theta$ and the angular momentum operator $J$ are paradigmatic of the complementary pairs of kinematic operators of all periodic quantum systems in that one member of this pair, the angular momentum operator, has a discrete spectrum and the other, the angle operator, has a periodic continuous spectrum. This difference in the spectra of the complementary operators has a profound significance for the nature of quantum models that describe periodic systems and contrasts with the operators for the complementary pair of observables position and momentum that both have continuous spectra and are unitary isomorphisms of each other.

 In this section we explore the action of $J$ and $\Theta$ on the Hilbert space $L^2 (\mathbb{T})$ of square integrable functions on the unit circle $\mathbb{T}$ and examine the critical  importance of the difference in their spectra for the Hilbert space representation of quantum state space.

\subsection{Hilbert space}
The set of eigenfunctions 
\[
    \bigg\{ e_k = \frac{1}{\sqrt{2\pi}}e^{i k \theta}\bigg\}_{k=-\infty}^\infty 
\]
is an orthonormal basis for the Hilbert space $L^2_{2\pi} (\mathbb{R})$ of $2\pi$-periodic square integrable functions on $\mathbb{R}$ that can be naturally identified with the Hilbert space $L^2 (\mathbb{T})$ of square integrable functions on the unit circle $\mathbb{T}$ via the mapping 
\begin{equation*}\label{eq2}
f(e^{it}) = F(t), \qquad f\in L^2(\mathbb{T}), \, F \in L^2_{2\pi} (\mathbb{R})\ \mathrm{and}\   t\in \mathbb{R}
\end{equation*}
that is based upon the canonical homomorphism of the real line into the complex plane $t\longrightarrow e^{it}$.\footnote{The Hilbert spaces $L^2 (\mathbb{T})$ and  $L^2_{2\pi} (\mathbb{R})$ can also be identified with the Hilbert space $L^2[-\pi,\pi]$ \cite{Rudin} . }  These Hilbert spaces are equipped with the inner product
\[
    (g,f)_{L^2(\mathbb{T})} =(G,F)_{L^2_{2\pi} (\mathbb{R})} = \int_{-\pi}^{\pi} G^\ast(\theta)F(\theta) d\theta 
\]
Thus for any $f\in L^2 (\mathbb{T})$ we can write the Fourier series

\begin{equation}\label{eq3}
    f = \sum_{k=-\infty}^\infty f_k e_k\,,
\end{equation}
where
\[
    f_k  = (f, e_k)_{L^2}\quad\mathrm{and}\quad \sum_{k=-\infty}^\infty \vert f_k\vert^2 <
    \infty\,
\]
that is, $\widehat{f}= \big\{f_k\big\}_{-\infty}^{\infty}\in l^2$ the Hilbert space of all square summable sequences.

\subsection{The angular momentum operator}

The self adjoint angular momentum operator is given in the angle representation for the planar rotator by the rule
\[
    J = - i \frac{d}{d \theta}
\]
and domain
\[
    D(J) = \Big\{ f\in L^2(\mathbb{T})\,:\, f^\prime\in L^2(\mathbb{T})\ \mathrm{and}\ f(-\pi) = f(\pi) \Big\}
\]
where we have set Planck's constant $\hbar = 1$ as we will do throughout the rest of the paper.
The basis elements $e_k$ of the Hilbert space $L^2(\mathbb{T})$ are eigenstates of the angular momentum operator $J$ with eigenvalues $k$\,:
\[
J\, e_k = -i\frac{d}{d\theta} \bigg(\frac{1}{\sqrt{2 \pi}} e^{i\,k \, \theta} \bigg) = k\, e_k \, 
\]
and $J$ admits the spectral decomposition
\begin{equation}\label{eq4}
J = \sum_{k = -\infty}^\infty k\, e_k  \, (e_k,^ \centerdot)
\end{equation}
where an intuitive notation for the functional operator $(e_k,^\centerdot) $ is used such that $(e_k,^\centerdot) f = (e_k,f)$ for $f \in L^2(\mathbb{T)}$ \cite{Akheizer}. 
This spectral decomposition is important in the theory of quantum mechanics because it enables us to work with functions of the unbounded operator $J$, for example $J$ generates a continuous unitary group on $L^2(\mathbb{T})$ with group elements $e^{-i\, y \, J}$. This can be proved by noting the spectral decomposition
\begin{equation} \label{eq5}
V_y := e^{-i\, y \, J} = \sum_{k = -\infty}^\infty e^{-i \, y \,k}\, e_k  \, (e_k,^ \centerdot), \qquad y \in [-\pi, \pi] \, ,
\end{equation}
from which it follows that
\[
e^{-i\, y \, J} \, e_k(x) = \frac{1}{\sqrt{2 \pi}} e^{i\,k \, x}\, e^{i\,k \, x} = e_k(x-y)
\]
and thus from equation (2) that
\[
e^{-i\, y \, J} \, f(x) = f(x-y)
\]
for an $f \in L^2(\mathbb{T})$.

\subsection{The angle operator}

The continuous bounded angle operator $\Theta$ is defined in the angle representation of the planar rotator by the multiplication rule
\[
    \Theta f(\theta) = \theta f(\theta)\,, \ \theta\in [-\pi,\pi]
\]
and a domain that spans the full Hilbert space
\[
    D(\Theta) = \Big\{ f\in L^2(\mathbb{T})\,:\, \theta f\in L^2(\mathbb{T}) \Big\}\,.
\]
This operator is self-adjoint and admits the spectral decomposition
\[
    \Theta  = \int_{-\pi}^{\pi} \theta E(d\theta)\,,
\]
in the Hilbert space  $L^2(\mathbb{T})$ where
\[
    E(B) = I_B(\theta)\quad\mathrm{for}\quad B\in \mathcal{A}\big([-\pi,\pi]\big)
\]
and $\mathcal{A}\big([-\pi,\pi]\big)$ is the $\sigma$-algebra of all Borel subsets of $[-\pi,\pi]$ \cite{Akheizer}.

More than this the operator $\Theta$ generates a discrete unitary group on $L^2(\mathbb{T})$ :
\begin{equation} \label{eq6}
   U_n := e^{i n\Theta} =  \sum_{k=-\infty}^\infty  e_{k+n}\,(e_k,\cdot)\,,\quad n\in \mathbb{Z}
\end{equation}
since
\[
     e^{i \Theta} e_k = \frac{1}{\sqrt{2\pi}}e^{i  x}\,e^{i k x} = \frac{1}{\sqrt{2\pi}} e^{i (k+1) x} = e_{k+1}
\]
that is $e^{i \Theta}$ is the angular momentum creation operator and similarly $e^{-i \Theta}$ is the angular momentum annihilation operator. 

However while the eigen-equation  for the operator $\Theta$
\begin{equation*}
    \big(\Theta - \theta I \big)f(x) = (x - \theta)f(x)\,, \quad x\in [-\pi,\pi]
\end{equation*}
does have the formal solutions $f_\theta(x)$ for every fixed $\theta\in [-\pi,\pi]$ these do not exist in the Hilbert space $L^2(\mathbb{T})$ a fact that is intimately related to the continuous spectrum $\sigma(\Theta) = [-\pi,\pi]$  \cite{Akheizer}.
In order to admit these formal solutions to a quantum theory we need to extend its state space beyond that of the Hilbert space $L^2(\mathbb{T})$.

\section{A rigged Hilbert state space}
In a search for a larger state space that contains the eigenstates of operators with both continuous and discrete spectra we note that in Hilbert space quantum theory the probability amplitude between the states $\psi,\chi\in L^2(\mathbb{T})$ is given by the inner product $(\psi,\chi)$ of these states. This provides a hint that a possible way forward is to use a dual structure for the state space in which we consider the linear functionals $(\psi,^.)$ with densities $\psi\in L^2(\mathbb{T})$ as one component and the states $\chi \in L^2(\mathbb{T})$ as their dual. It is then a short step to follow Schwartz \cite{Schwartz} and consider the space $\mathcal{D}^\prime (\mathbb{T})$ of all bounded linear functionals (distributions) on the unit circle that act upon the space $C^\infty(\mathbb{T})$ of infinitely differentiable functions. In emulation of the symbol $(\,, )$ for the inner product we denote the dual pairing $\langle F,\phi \rangle \in \mathbb{C}$ between $F \in \mathcal{D}^\prime(\mathbb{T)}$ and $\phi \in C^\infty(\mathbb{T})$ by the symbol $\langle \, , \rangle$ \cite{Dautray}. As an example of the use of this notation we define $\delta_\theta \in \mathcal{D}^\prime (\mathbb{T})$ as the Dirac measure on $\mathbb{T}$ by the identification  
\[
\langle \delta_\theta,\phi\rangle := \phi(\theta).
\]
Thus we can calculate the Fourier coefficient of order  $k$ of the distribution $F\in \mathcal{D}^\prime (\mathbb{T})$ as
\[
F_k = \langle F, \frac{1}{\sqrt{2\pi}}e^{-ikx} \rangle \in \mathbb{C},\quad x\in[-\pi,\pi ]\, ,
\]
this is well defined since $e^{-ikx} \in C^\infty(\mathbb{T})$ for any $k \in\mathbb{Z}$. Hence we obtain the Fourier series
\[
F= \sum^\infty_{k=-\infty} F_k \,e_k = \frac{1}{\sqrt{2\pi}}\sum^\infty_{k=-\infty} F_k \,e^{ikx}
\]
that converges in the space $\mathcal{D}^\prime (\mathbb{T})$ to the distribution $F$, that is the Fourier coefficients are of slow growth, in particular
\[
 \lim_{ |k| \to \infty} \frac{|F_k|}{(1+k^2)^j} = 0
 \]
for some sufficiently large $j \in \mathbb{R}$ and 
\begin{equation*} \label{modgs} 
\hat{F} = \Big \{F_k \Big \}_{k = - \infty}^\infty \in s^\prime
\end{equation*}
the set of all sequences of moderate growth \cite{Dautray}. For example the Fourier coefficients of the Dirac measure $\delta_\theta$ are 
\[
[\delta_\theta]_k = \langle \delta_\theta , e^{ikx} \rangle =\frac{1}{\sqrt{2\pi}} e^{ik\theta} 
\]
and hence its Fourier series is
\[
\delta_\theta (x) =  \frac{1}{\sqrt{2\pi}}\sum^\infty_{k=-\infty} e^{ik\theta}e^{ikx}\, .
\]

Further insight into the nature of the periodic Dirac measure $\delta_0$ can be obtained by relating it to the the Dirac measure on the real line $\hat{\delta}_0 \in \mathcal{D}^\prime (\mathbb{R})$, with $\langle \hat{\delta}_0,f\rangle := f(0)$ for $f\in C^\infty(\mathbb{R})$, via the equation
\begin{equation} \label{delta}
\delta_0(x) = \sum^\infty_{k = -\infty} \hat{\delta}_0(x+ k \, 2\pi)
\end{equation}
that gives credence to the notion that the periodic Dirac measure $\delta_0$ can be considered as an equally spaced comb of the Dirac measures $\hat{\delta}_0$ \cite{Dautray}. 

\subsection{Generalised eigenfunctions and rigged Hilbert space}
We note that each $\phi \in C^\infty(\mathbb{T})$ has Fourier coefficients $\phi_k$ such that the set $\big \{ \phi_k \big \}^\infty_{-\infty} \in s$ the set of all rapidly decreasing sequences and that there exists a Gelfand triple 
\[
s\subset \ell^2 \subset s^\prime
\]
that denotes the rigging of the Hilbert space $\ell^2$ by $s$ and its dual $s^\prime$ \cite{Dautray}. Parseval's relation and the linearity of the Fourier relations implies that this rigged Hilbert space of sequences is isomorphic to the rigged Hilbert space
\[
C^\infty(\mathbb{T}) \subset L^2(\mathbb{T}) \subset \mathcal{D}^\prime (\mathbb{T})
\]
on the unit circle. We consider the action of the angle operator $\Theta$ and angular momentum operator $J$ on this rigged Hilbert space and note that
\[
C^\infty(\mathbb{T}) \subseteq D(\Theta^n ), D(J^n)\,\, \mathrm{for\,\, any}\,\, n\in \mathbb{N}
\]
therefore we can define $\Theta$ on $\mathcal{D}^\prime  (\mathbb{T})$ by the dual pair relation
\[
    \langle \Theta F,\varphi \rangle = \langle  F, \Theta\varphi \rangle \,,\qquad F\in \mathcal{D}^\prime  (\mathbb{T})\,,\ \varphi\in \mathcal{C}^\infty  (\mathbb{T})\,.
\]
In particular, for the Dirac measure $\delta_\theta$ we have
\[
    \langle \Theta \delta_\theta,\varphi \rangle = \langle  \delta_\theta, \Theta\varphi \rangle = \langle  \delta_\theta, x\varphi(x) \rangle = \theta\varphi(\theta) = \theta\langle  \delta_\theta, \varphi \rangle = \langle  \theta\delta_\theta, \varphi \rangle\,,
\]
thus
\[
    \Theta \delta_\theta = \theta \delta_\theta\,
\]
that is, the distribution $\delta_\theta$ is a generalised eigenfunction for the operator $\Theta$ that is contained in $\mathcal{D}^\prime (\mathbb{T})$.

We show in an appendix that these eigenfunctions are orthogonal in the sense that
\[
\langle \delta_\theta , \delta_{\theta^\prime} \rangle_g = \delta_{(\theta - \theta^\prime)}
\]
where  $\langle  ,  \rangle_g$ is a generalised sesquilinear form and thus implement one of Dirac's heuristic ideas.

 \subsection{Generalised eigenstate decompositions of operators with continuous spectra} 
Borel functions of the self adjoint angle operator have generalised eigenstate decompositions on the rigged Hilbert space $C^\infty(\mathbb{T}) \subset L^2(\mathbb{T}) \subset D^\prime (\mathbb{T})$ that parallel the eigenstate decomposition in the Hilbert space $L^2(\mathbb{T})$ of the angular momentum operator that was given in equation (\ref{eq4}).\footnote{We note that generalised eigenstate decompositions do not exist for all operators with continuous spectra.} In order to prove this consider the generalised projection operator
\[
\Pi(\theta):=\delta_\theta\langle\delta_\theta,^.\rangle^*\,:\,\mathcal{C}^\infty(\mathbb{T})\rightarrow\mathcal{D}^\prime(\mathbb{T}),
\]
that is for $\phi\in\mathcal{C}^\infty(\mathbb{T})$
\[
\Pi(\theta)\phi=\langle\delta_\theta,\phi\rangle^*\delta_\theta=\phi^*(\theta)\delta_\theta\in\mathcal{D}^\prime(\mathbb{T})
\]
and therefore for $\psi\in\mathcal{C}^\infty(\mathbb{T})$ we have
\[
\langle\Pi(\theta)\phi,\psi\rangle = \langle\phi^*(\theta)\delta_\theta,\psi\rangle = \phi^*(\theta)\psi(\theta).
\]
This generalised projection operator $\Pi(\theta)$ is an example of an eigenoperator as introduced by Roberts \cite{Roberts}, \cite{Melsheimer1}, \cite{Melsheimer2} in his studies of quantum theory in rigged Hilbert space. The results obtained from these well defined mappings from the space of infinitely differentiable functions to the space of distributions enhance the weak notions of spectral decomposition that are used by other authors \cite{Hytonen}. 

By using these eigenstate projection operators as primitives generalised decompositions of operators that can be expressed as Borel functions $f$ of the angle operate can be obtained. This can be proved by considering the generalised spectral decomposition 
\[
f(\tilde{\Theta}) = \int_{-\pi}^{\pi}d\theta\,f(\theta)\,\Pi(\theta) = \int_{-\pi}^{\pi}d\theta\,f(\theta)\,\delta_\theta\langle\delta_\theta,^.\rangle^*\,:\,\mathcal{C}^\infty(\mathbb{T})\rightarrow\mathcal{D}^\prime(\mathbb{T}),
\]
where the tilde denotes the generalised status of the operator. For each $\phi\in\mathcal{C}^\infty(\mathbb{T})$ the distribution
\[
F_\phi = \int_{-\pi}^{\pi}d\theta\,f(\theta)\,\phi^*(\theta)\delta_\theta \in\mathcal{D}^\prime(\mathbb{T})
\]
is well-defined as
\[
\langle F_\phi, \psi\rangle := \int_{-\pi}^{\pi}d\theta\,\langle f(\theta)\,\phi^*(\theta)\delta_\theta,\psi\rangle = \int_{-\pi}^{\pi}f(\theta)\,\phi^*(\theta)\,\psi(\theta)\,d\theta \in\mathcal{D}^\prime(\mathbb{T}).
\]

Hence the powers  $\tilde{\Theta}^n$ of the angle generalised operators have the generalised eigenstate decompositions 
\begin{equation} \label{moms}
\tilde{\Theta}^n = \int_{-\pi}^{\pi}d\theta\,\theta^n\,\Pi(\theta) = \int_{-\pi}^{\pi}d\theta\,\theta^n\,\delta_\theta\langle\delta_\theta,^.\rangle^*\,:\,\mathcal{C}^\infty(\mathbb{T})\rightarrow\mathcal{D}^\prime(\mathbb{T}).
\end{equation}
 In particular for $n = 0$ we have the eigenstate decomposition of the unit generalised operator
\[
\tilde{I} = \int_{-\pi}^{\pi}d\theta\,\Pi(\theta) = \int_{-\pi}^{\pi}d\theta\,\delta_\theta\langle\delta_\theta,^.\rangle^*\,:\,\mathcal{C}^\infty(\mathbb{T})\rightarrow\mathcal{D}^\prime(\mathbb{T}),
\]
that provides a rigorous justification of the Dirac expression 
\[
I=\int_{-\pi}^{\pi}d\theta\,\vert \theta\rangle\langle \theta\vert 
\]
and a basis for a statistical interpretation of a quantum theory of the planar rotator with respect to the angle.  

Another important generalised operator with an angle spectral decomposition is the angular momentum group displacement operator of equation (\ref{eq6})
\[
\tilde{U}_n:=e^{i n \tilde{\Theta}}
\]
with
\[
e^{i n \tilde{\Theta}} = \int_{-\pi}^{\pi}d\theta\, e^{i n\theta}\,\Pi(\theta) = \int_{-\pi}^{\pi}d\theta\,e^{i n\theta}\,\delta_\theta\langle\delta_\theta,^.\rangle^*\,:\,\mathcal{C}^\infty(\mathbb{T})\rightarrow\mathcal{D}^\prime(\mathbb{T}),
\]
and thus
\[
\langle e^{i n\tilde{\Theta}}\phi,\phi\rangle = \int_{-\pi}^{\pi} d\theta\, e^{i n\theta}\,\langle\delta_\theta,\phi\rangle^*\,\langle\delta_\theta,\phi\rangle =\int_{-\pi}^{\pi} e^{i n\theta}\,\phi^*(\theta)\,\phi(\theta) d\theta.
\]

The form of the spectral decompositions of equations (\ref{eq4}) and (\ref{moms}) show that the use of  a rigged Hilbert state space provides a rigorous quantum theory that has a deep inherent symmetry in its mathematical representation of operators with both continuous and discrete spectra.

\section{Minimum uncertainty states and Colombeau algebra}
We now turn from the analysis of functions of individual operators operators to the more profound subject of compound operators that are constructed by multiplying functions of the operators of complementary variables. In particular we note that the quantum nature of the planar rotator model is manifest by the Weyl commutation relation
\begin{equation} \label{eq8}
U_n V_y = e^{iyn} V_y U_n
\end{equation} 
for the discrete unitary group elements $U_n = e^{i n\Theta}$ and continuous unitary group elements $V_y = e^{ -i y J}$ on  $L^2(\mathbb{T})$. This commutation relation applies to the complete Hilbert space $L^2(\mathbb{T})$ and can be derived by applying the operator expansions of equation (\ref{eq5}) and equation (\ref{eq6}) to obtain
\[
U_n V_y = \sum_{k=-\infty}^\infty e^{-iyk} e_{k+n}\,(e_k,\cdot) = e^{iyn}\sum_{k=-\infty}^\infty e^{-iy(k+n)} e_{k+n}\,(e_k,\cdot) = e^{iyn}\,V_yU_n\,.
\]
Since the group operators $U_n$ and $V_y$ do not commute the angle and angular momentum variables cannot be simultaneously measured with perfect precision because if a quantum state $\psi$ existed with a precise angle $\theta$ and angular momentum $k$ then $U_n V_y\psi = e^{in\theta} e^{iyk} \psi$ and $ V_yU_n\psi = e^{iyk} e^{in\theta} \psi$ that implies $U_n V_y - V_y U_n = 0$ in contradiction to the commutation relation given in equation (\ref{eq8}).

The inability to make a simultaneous precise measurement of the angle and angular momentum raises the question of whether states exist that minimise the uncertainty product $\Delta \Theta \, \Delta J$ where $\Delta \Theta$ is a measure of uncertainty of the angle and $\Delta J$ is a measure of uncertainty of the angular momentum. It has been proved that minimum variance functions for the planar rotator do not exist in the Hilbert space $L^2(\mathbb{T})$ \cite{Holevo}. Hence our search for minimum variance states is focussed on whether generalised functions exist that minimise the uncertainty product $\Delta \Theta \, \Delta J$ for the planar rotator. We begin to ask this question by investigating the possibility of suitable measures of uncertainty $\Delta \Theta$ and $\Delta J$ and associated expectations of the angle and angular momentum variable.

\subsection{Measures of uncertainty}
Some care needs to be taken in defining the expectations of the moments of a random variable on the unit circle like the angle $\theta$, for example the periodicity of the angle variable can lead to an ambiguity in the angular mean of a given probability distribution. A precise definition of the mean angle (direction) is $\bar{\Theta} = \mathrm{arg}(I)$ where
\[
I = \int_\mathbb{T}e^{i \theta} P(\theta) d\theta
\]
for the probability density $P(\theta)$ and integration over an interval covering the unit circle $\mathbb{T}$ \cite{Mardia}.

In order to assist in the search for minimum uncertainty functions we consider the states $\psi_{0,0}(\theta)$ that have probability densities $P_{0,0} (\theta)= \vert \psi_{0,0} (\theta) \vert ^2$ with zero mean angle $\bar{\Theta}=0$ and zero mean angular momentum $\bar{J}=0$. A state
\[
 \psi_{\bar{\Theta},0}(\theta) = V_{\bar{\Theta}}\,\psi_{0,0}(\theta)
\]
 with arbitrary mean angle $\bar{\Theta}$ can be obtained from this state by applying the angle shift operator $V_{\bar{\Theta}}$ of equation (\ref{eq5}). More than this it can be seen from the angular momentum operator rule $J = i \frac{d}{d\theta}$ and the differential product rule that the state
\begin{equation} \label{usf}
  \psi_{\bar{\Theta},\bar{J}}(\theta) = e^{i\bar{J}\theta}\,\psi_{\bar{\Theta},0}(\theta) = e^{i\bar{J}\theta}\,V_{\bar{\Theta}}\,\psi_{0,0}(\theta)
\end{equation}
has mean angular momentum $\bar{J}$ and mean angle $\bar{\Theta}$. Hence we can limit our search for minimum uncertainty states to ones with zero mean angle and zero mean angular momentum and generate a family of minimum uncertainty states with arbitrary means from them.
 
In anticipation that the probability density for the angle representation of a minimum uncertainty state will be concentrated about its mean we have taken particular care to place that mean in the centre of the interval $[-\pi,\pi]$ so that we can define the uncertainty $\Delta \theta$ as the positive square root of the variance 
\begin{equation} \label{eq9}
(\Delta \Theta)^2 = \mathrm{Var(\Theta)} = (\psi, \Theta^2 \psi) = (\Theta \psi,\Theta \psi)
\end{equation}
on the interval $[-\pi,\pi]$ rather than as a circular measure \cite{Mardia}, and the final equality follows from the self adjoint property of $\Theta$. \footnote{We note that this definition of the uncertainty of a circular random variable that is concentrated about its mean is supported by arguments that are given by Mardia et. al. \cite{Mardia1}, p.427} This choice of uncertainty enables us to directly compare the periodic case with other states on the real line  and also provides a solution for our problem in terms of a simple differential equation \cite{Merzbacher}. 
 
The measure of uncertainty $\Delta J$ for the angular momentum is chosen to be the positive square root of the variance 
\begin{equation} \label{eq10}
(\Delta J)^2 = \mathrm{Var(J)} = (\psi, J^2 \psi) = (J \psi, J \psi).
\end{equation}
\subsection{Minimum uncertainty states}
Given the definitions for the variances of $\Theta$ and $J$ that are given in equations (\ref{eq9}) and (\ref{eq10}) respectively a generalised Schwartz inequality 
\[
(\Delta \Theta)^2\,(\Delta J)^2 \geqslant \, \vert (\psi, \Theta \, J \psi ) \vert ^2
\]
for for the states $\Theta \psi$ and $J \psi$ can be obtained  \cite{Merzbacher}, where the equality holds if and only if 
\[ 
J \psi = \lambda \, \Theta \psi  
\]
for $\lambda \in \mathbb{C}$. By substituting with the rules $J = -i \frac{d}{d \theta}$ and multiplication by $\theta$ for the operator $\Theta$ into this equation we obtain the differential equation
\begin{equation} \label{mudec}
\bigg(\frac{d}{d \theta} + \lambda \, \theta \bigg) \psi (\theta ) = 0 \, .
\end{equation}

 It is well known that the Gaussian functions
\[
g_\varepsilon(x) = \frac{1}{\sqrt{\pi\varepsilon}} e^{- x^2/\varepsilon}\,,\quad x\in \mathbb{R}\,,
\]
satisfy the analogous equation
\[
\bigg(\frac{d}{dx} + \lambda \, x\bigg) g(x) = 0
\]
on the real line $\mathbb{R}$ with $\lambda = 2/\varepsilon$ \cite{Merzbacher}. Therefore in seeking a solution to equation (\ref{mudec}) we consider the periodic functions
\begin{equation*} \label{1par}
   \psi_\varepsilon(\theta) = \sum_{k=-\infty}^{\infty} g_\varepsilon(\theta + 2\pi k),
\end{equation*}
that are parameterised by $\varepsilon$, to which we apply the operator $\frac{d}{d\theta} + \frac{2}{\varepsilon} \theta$ and obtain
\begin{eqnarray}\label{remdr}
  \Psi_\varepsilon(\theta) &:=& \bigg( \frac{d}{d\theta} + \frac{2}{\varepsilon} \theta\bigg) \psi_\varepsilon(\theta) = \sum_{k=-\infty}^{\infty} \bigg( \frac{d}{d\theta} + \frac{2}{\varepsilon} \theta\bigg) g_\varepsilon(\theta + 2\pi k)\\ \nonumber \\ &=& \frac{1}{\sqrt{\pi\varepsilon}}\sum_{k=-\infty}^{\infty} \bigg(- \frac{2}{\varepsilon} \theta -\frac{4}{\varepsilon}\pi k + \frac{2}{\varepsilon} \theta \bigg) e^{-  (\theta + 2\pi k)^2/\varepsilon}\nonumber\\ &=&
     - \frac{4\sqrt{\pi}}{\varepsilon\sqrt{\varepsilon}}\sum_{k=-\infty}^{\infty} k\,e^{- (\theta + 2\pi k)^2/\varepsilon}\,.\nonumber
\end{eqnarray}
Hence the magnitude $ \big\vert \Psi_\varepsilon(\theta)\big\vert $ is bounded above by the estimate
\begin{eqnarray*}
  \big\vert \Psi_\varepsilon(\theta)\big\vert \leq  \frac{4\sqrt{\pi}}{\varepsilon\sqrt{\varepsilon}}\bigg(\Big\vert\sum_{k=-\infty}^{-1} k\,e^{- (\theta + 2\pi k)^2/\varepsilon}\Big\vert\, + \sum_{k=1}^{\infty} \Big\vert  k\, e^{- (\theta + 2\pi k)^2/\varepsilon}\Big\vert \bigg) ,
\end{eqnarray*}
and on the interval $\theta\in [-\pi,\pi]$
\begin{eqnarray*}
  \big\vert \Psi_\varepsilon(\theta)\big\vert \leq  \frac{8\sqrt{\pi}}{\varepsilon\sqrt{\varepsilon}}\sum_{k=1}^{\infty} k\,e^{- \pi^2 (2k -1)^2/\varepsilon} = \frac{8\sqrt{\pi}}{\varepsilon\sqrt{\varepsilon}}\, e^{- \pi^2/\varepsilon}\sum_{k=1}^{\infty}k\,\big(e^{- \pi^2/\varepsilon}\big)^{4k(k -1)} ,
\end{eqnarray*}
where the first few terms in the final sum are given explicitly in the square brackets of the relation
\begin{equation*} \label{bound}
  \big\vert \Psi_\varepsilon(\theta)\big\vert \leq  \frac{8\sqrt{\pi}}{\varepsilon\sqrt{\varepsilon}}\, e^{- \pi^2/\varepsilon}\bigg[ 1+2\big(e^{- \pi^2/\varepsilon}\big)^8+3\big(e^{- \pi^2/\varepsilon}\big)^{24}\,...\,...\,\bigg],
\end{equation*}
that implies 
\[
    \lim_{\varepsilon\to 0} \Psi_\varepsilon(\theta) = 0\,.
\]
Thus a minimum variance generalized function exists as a limit of the sequence of functions $ \psi_\varepsilon(\theta) $ that satisfies equation  (\ref{mudec}). It is tempting to identify this limit as a periodic Dirac measure. However the Dirac measure
\[
    \delta_0(\theta) = \sum_{k=-\infty}^\infty \hat{\delta}_0(\theta +2\pi k )\in \mathcal{D}^\prime (\mathbb{T})\,,
\]
of equation (\ref{delta}), with $\hat{\delta}_0 \in \mathcal{D}^\prime (\mathbb{R})$, does not satisfy equation (\ref{mudec}) since $\theta \, \delta_0 (\theta) = 0$ in  $\mathcal{D}^\prime (\mathbb{T})$ and ${\delta_0}^\prime (\theta) \neq 0$. This failure of the theory of distributions to provide a solution for the differential equation (\ref{mudec}) with non constant coefficients is intrinsic to the theory and follows from the fact that a product of distributions is not always a distribution, as proved by Schwartz in his impossibility theorem \cite{SchwartzI}. In order to handle terms like $\theta\, \delta (\theta)$ in the minimum uncertainty differential equation (\ref{mudec}) we need an algebra of generalised functions not just a vector space. While there are many ways to provide such an algebra a simple and natural generalisation of the space of distributions is provided by the special algebra $\mathcal{G}(\mathbb{T})$ of Jean-Francis Colombeau \cite{Colombeau1, Colombeau2}, into which the space of distributions $D^\prime (\mathbb{T})$ can be canonically embedded \cite{Prazak}.
 
 \subsection{Colombeau generalised functions}
 In the following section we first give a  concise derivation of a special Colombeau algebra, a helpful primer for Colombeau algebras can be found in \cite{Gsponer} while a recent comprehensive account is given in \cite{Grosser}. We then apply the theory of Colombeau algebras to a proof that minimum uncertainty states exist as Colombeau generalised functions. Finally we demonstrate the association between these minimum uncertainty states and the Dirac measure as a distribution.
 
 \subsubsection{Derivation}
 We can derive the space $\mathcal{G}(\mathbb{T})$ that is an associative, commutative differential algebra with pointwise operations by considering the space $\mathcal{E}(\mathbb{T})$ of all \textit{one-parameter families} (or \textit{net}) of infinitely differentiable functions from   $\mathcal{C}^\infty (\mathbb{T})$:
\[
    \mathcal{E}(\mathbb{T}) := \Big\{ (u_\varepsilon) = \big\{ u_\varepsilon(\cdot)\in \mathcal{C}^\infty (\mathbb{T}),\ \varepsilon\in(0,1]\big\} \Big\}\,,
\]
that is exemplified by the periodic functions
\begin{equation} \label{net}
   \psi_\varepsilon(\theta) = \sum_{k=-\infty}^{\infty} g_\varepsilon(\theta + 2\pi k),\, \, \varepsilon\in(0,1]
\end{equation}
derived from equation (\ref{1par}). 
Then
\begin{eqnarray*}
	    \lefteqn{\mathcal{E}_M (\mathbb{T}) := \bigg\{ (u_\varepsilon)\in \mathcal{E}(\mathbb{T})\,:\, \forall j\in \mathbb{N}_0\ \exists q\in \mathbb{N}\quad}\\
	    &&\qquad\qquad\qquad\qquad\qquad\qquad \sup_{x\in[-\pi,\pi]} \bigg\vert \frac{d^j}{dx^j} u_\varepsilon(x)\bigg\vert  = \mathcal{O}(\varepsilon^{-q})\quad\mathrm{as}\ \varepsilon\to 0\bigg\}
\end{eqnarray*}
is the space of all nets of \textit{moderate growth}  and
\[
     \mathcal{N} (\mathbb{T}) := \bigg\{ (u_\varepsilon)\in \mathcal{E}_M(\mathbb{T})\,:\, \forall q\in \mathbb{N}\quad \sup_{x\in[-\pi,\pi]} \big\vert u_\varepsilon(x)\big\vert  = \mathcal{O}(\varepsilon^{q})\quad\mathrm{as}\ \varepsilon\to 0\bigg\}
\]
is the subspace of the \textit{negligible elements} of $\mathcal{E}_M (\mathbb{T})$. The space $\mathcal{N} (\mathbb{T})$ is an ideal in $\mathcal{E}_M (\mathbb{T})$; and the special Colombeau algebra of periodic generalised functions is defined as the quotient
\[
    \mathcal{G}(\mathbb{T}):= \mathcal{E}_M (\mathbb{T})/\mathcal{N} (\mathbb{T})\,.
\]
 The elements $u\in \mathcal{G}(\mathbb{T})$ are the equivalence classes $u = \big[ (u_\varepsilon)\big]$ in analogy to the elements of Hilbert space \cite{Rudin} and we say that $(u_\varepsilon)\in \mathcal{E}_M (\mathbb{T})$ is a representative of $u$ \cite{Colombeau2, Grosser}.  
 
 Finally, we introduce the ring of \textit{generalised numbers} $\mathcal{K} := \mathcal{E}_M/\mathcal{N}$, where
\begin{eqnarray*}
  \mathcal{E}_M &=&  \Big\{ (r_\varepsilon)\in \mathcal{E}\,:\, \exists q\in \mathbb{N}\quad \vert  r_\varepsilon\vert  = \mathcal{O}(\varepsilon^{-q})\quad\mathrm{as}\ \varepsilon\to 0\Big\}\,,\\
  \mathcal{N} &=&  \Big\{ (r_\varepsilon)\in \mathcal{E}_M\,:\, \forall q\in \mathbb{N}\quad \vert r_\varepsilon\vert  = \mathcal{O}(\varepsilon^{q})\quad\mathrm{as}\ \varepsilon\to 0\Big\}\,,\\
  \mathcal{E} &=& \Big\{ (r_\varepsilon) = \big\{ r_\varepsilon\in \mathbb{K},\ \varepsilon\in(0,1]\big\} \Big\}
\end{eqnarray*}
and $\mathbb{K}$ is either $\mathbb{R}$ or $\mathbb{C}$. We follow the custom of writing $\mathcal{R}$ for generalised real numbers and $\mathcal{C}$ for generalised complex numbers.
 
 \subsubsection{Minimum uncertainty states revisited}
We note that according to the estimate (\ref{bound}), the generalised function $\Psi = [(\Psi_\varepsilon)]$ defined in equation (\ref{remdr}) satisfies the condition
\[
     \forall q\in \mathbb{N}\quad \sup_{x\in[-\pi,\pi]} \big\vert \Psi_\varepsilon(x)\big\vert  = \mathcal{O}(\varepsilon^{q})\quad\mathrm{as}\ \varepsilon\to 0\,,
\]
that is $\Psi \in \mathcal{N}(\mathbb{T})$ and therefore $\Psi = 0$ in $\mathcal{G}(\mathbb{T})$. This implies that the generalised function $\psi = [(\psi_\varepsilon)]\in \mathcal{G}(\mathbb{T})$ that is defined by equation (\ref{net}) satisfies equation (\ref{mudec}) in the space $\mathcal{G}(\mathbb{T})$:
\[
    \psi^\prime + \lambda x \psi = 0\,,
\]
where $x = [(x_\varepsilon)]\in \mathcal{G}(\mathbb{T})$ with $x_\varepsilon \equiv x$ and $\lambda = [(\lambda_\varepsilon)]\in \mathcal{R}$ the generalised real numbers with $\lambda_\varepsilon = \varepsilon^{-1}$. 

That is a minimum uncertainty state $\psi$ of the planar rotator with zero mean angle and angular momentum exists in the Colombeau space $\mathcal{G}(\mathbb{T})$ and with a promotion of the operators $e^{i\bar{J} \theta}$ and $V_{\bar{\Theta}}$ of equation (\ref{usf}) to the Colombeau algebra this gives rise to the family of minimum uncertainty states
\[
\psi_{\bar{\Theta},\bar{J}}= e^{i\bar{J}\theta}\,V_{\bar{\Theta}}\,\psi 
\]
where $\psi_{\bar{\Theta},\bar{J}} \in \mathcal{G}(\mathbb{T})$.

\subsubsection{Distributions}
One of the questions that is raised by the existence of the minimum uncertainty state $\psi \in \mathcal{G}(\mathbb{T})$ and its formulation in terms of a net of infinitely differentiable functions is whether a relationship exists between it and the Dirac measure $\delta \in D^\prime (\mathbb{T})$. 

In order to answer this question we note that the space $\mathcal{C}^\infty (\mathbb{T})$ is a subalgebra of $\mathcal{G}(\mathbb{T})$ with the constant embedding
\[
    f\in \mathcal{C}^\infty(\mathbb{T}) \mapsto (f_\varepsilon) + \mathcal{N} (\mathbb{T})\,,\quad \mathrm{where}\ f_\varepsilon\equiv f
\]
and that the space $\mathcal{D}^\prime (\mathbb{T})$ of distributions on the unit circle can be canonically embedded into the special Colombeau algebra of generalised functions on the unit circle: $\mathcal{D}^\prime (\mathbb{T}) \hookrightarrow \mathcal{G}(\mathbb{T})$ (see, for example, \cite{Valmorin}, \cite{Grosser}, \cite{Prazak}). This imbedding implies the existence of the spatial quintet
\[
    \mathcal{N} (\mathbb{T}) \subset \mathcal{C}^\infty (\mathbb{T})\subset \mathcal{L}^2 (\mathbb{T})\subset \mathcal{D}^\prime (\mathbb{T})\subset \mathcal{G}(\mathbb{T})
\]
and is exemplified by the Dirac measure 
\[
    \delta \in D^\prime(\mathbb{T}) \mapsto (\psi_\varepsilon) + \mathcal{N} (\mathbb{T})\quad\mathrm{with}\ \psi_\varepsilon(x) = \sum_{k=-\infty}^{\infty} g_\varepsilon(x + 2\pi k)\,.
\]
More than this we say that a generalised function $u\in \mathcal{G}(\mathbb{T})$ is \textit{associated} with a distribution $U\in \mathcal{D}^\prime (\mathbb{T})$ if
\[
    \lim_{\varepsilon\to 0} \int_{-\pi}^{\pi} u_\varepsilon (x) \varphi (x) dx = \langle U, \varphi\rangle\quad\mathrm{for\ all}\ \varphi\in \mathcal{C}^\infty (\mathbb{T})\,,
\]
and we write $u \approx U$. Hence the generalised function $\psi = [(\psi_\varepsilon)]$ that represents the minimum variance state of the planar rotator is associated with the Dirac measure on the unit circle $\delta$: $\psi\approx\delta$. However the inability of the Dirac measure on the unit circle to preserve nonlinear information disqualifies it as a solution of equation (\ref{mudec}) \cite{Grosser}.

\section{Discussion}
It has been demonstrated that the state space for periodic quantum theory must be expanded beyond that of Hilbert space and indeed that of rigged Hilbert space to a Colombeau algebra in order to represent key elements of periodic quantum systems such as minimum uncertainty states. The proof that these states exist for the simplest of all periodic quantum systems exemplifies the importance of Colombeau algebras in providing solutions for basic equations of quantum physics. 

Our approach to a Colombeau state space commenced with a Hilbert space that provided a prototypical analysis of quantum operators via the angular momentum operator with discrete spectra. Having noted that the continuous spectral angle operator did not have solutions in Hilbert space we calculated the Fourier coefficients for the Dirac measure on the unit circle. These coefficients are elements of the space of sequences with slow growth, that is a dual of the space of sequences of rapid decay. It was shown that these two spaces rig the Hilbert space of  square summable sequences. Parseval's relation and the linearity of the Fourier relations then enabled us to establish that the Hilbert space of square integrable functions on the unit circle can be rigged with the space of infinitely differentiable functions on the unit circle and its dual the space of linear continuous functionals that are defined on them. This rigged Hilbert space facilitated a quick calculation that showed that the Dirac measure $\delta_\theta$ on the unit circle is a simple representation of the eigenstates of the angle operator with continuous periodic spectrum. In order to further justify the assignment of these distributions to a state space, but not break the flow of the paper, we have provided in an appendix a rigorous demonstration that their generalised sesquilinear product is a distribution thus implementing one of Dirac's ideas.

In hindsight we have been able to see that the Schwartz distributions can be interpreted as a representation of the linear component of the special Colombeau algebra on the unit circle $\mathcal{G}(\mathbb{T})$ indeed it is linearly imbedded in this algebra. In order to illustrate the importance of the distributions for linear problems we proved the new result that in the rigged Hilbert space on the unit circle there exist generalised decompositions of operators that are formed as Borel functions of the angle operator with continuous periodic spectrum. These operators included the unit operator, with particular significance for the probabilistic interpretation of quantum mechanics, and the angular momentum creation and annihilation operators.

We then turned our attention to the profound topic of noncommutativity of the angle displacement group elements and the shift operators for angular momentum. In particular we examined the question of whether minimum uncertainty states exist for the simplest periodic quantum system via a differential equation with nonconstant coefficients. In order to solve this equation we had to expand the topology of the state space to that of the Colombeau algebra on the unit circle. Anticipating questions about the relationship between these Colombeau generalised functions and the Dirac measures we indicated that the space of distributions on the unit circle can be linearly embedded in the Colombeau algebra and that a notion of association exists between the minimum uncertainty states of planar rotator and the Dirac measures on the unit circle. This association indicates that the linear Dirac measure on the unit circle equates with the zero parameter limit of a nonlinearity preserving net of infinitely differentiable periodic Gaussian functions.

More than this we showed by a direct calculation of a limiting sequence of functions that solved the minimum uncertainty state equation that working with Colombeau algebras on the unit circle is as simple as working with an algebra of infinitely differentiable functions.

\section*{Appendix: Generalised sesquilinear forms}
The rigged Hilbert space $\mathcal{C}^\infty (\mathbb{T})\subset \mathcal{L}^2 (\mathbb{T})\subset \mathcal{D}^\prime (\mathbb{T})$ provides a vector space that contains all the eigenstates of the kinetic operators for the planar rotator. On this space we have defined and discussed the inner product $(f,g)_{L^2}$ and the dual pair relation  $\langle F,\psi \rangle$ between the periodic distributions $F \in \mathcal{D}^\prime (\mathbb{T})$ and the periodic test functions $\psi \in C^\infty(\mathbb{T})$. In order to complete the development of this state space for a quantum theory that utilises periodic generalised functions such as $\delta_\theta$ we need to develop a rigorous description of a generalised sesquilinear form $\langle F, G\rangle_g$ with variables $F,G\in \mathcal{D}^\prime (\mathbb{T})$.

To that end consider distributions $F,G\in \mathcal{D}^\prime (\mathbb{T})$ with  expansions
\[
     F (x) = \frac{1}{\sqrt{2\pi}}\sum_{n=-\infty}^{\infty} F_n\,e^{inx} = \sum_{n=-\infty}^{\infty} F_n\,e_n
\]
and
\[
     G(x) = \frac{1}{\sqrt{2\pi}}\sum_{m=-\infty}^{\infty} G_m\,e^{imx} = \sum_{m=-\infty}^{\infty} G_m\,e_m\,.
\]
We note that the existence of the generalised functions $F$ and $G$ and the rotational symmetry of the unit circle $\mathbb{T}$ implies the existence of the classes of functions
\[
    \Big\{F_\theta (x) = F (x-\theta )= \sum_{n=-\infty}^{\infty} F_n\, e^{in\theta}\,e_n : \theta \in [-\pi,\pi ] \Big\}
\]
and
\[
     \Big\{ G_\phi (x) = G (x-\phi )= \sum_{m=-\infty}^{\infty} G_m\,e^{im\phi}\,e_m : \phi \in [-\pi,\pi ] \Big\}
\]
 of which we understand $F$ and $G$ to be representative.
 
Since the functions
\[
     G^N(x)= \sum_{m=-N}^{N} G_m\,e_m
\]
are infinitely differentiable, we can construct the dual pairs
\[
     \langle F, G^N\rangle= \bigg\langle F, \sum_{m=-N}^N  G_m \,e_m\bigg\rangle = \sum_{n=-\infty}^\infty \sum_{m=-N}^N F^*_n G_m\, \langle e_n,  e_m\rangle\,,
\]
where we have used the sesquilinear nature of the dual pairing in writing the final term. The orthonormality of the functions $e_k$ enables us to write
\[
     \langle F, G^N\rangle=   \sum_{n=-N}^N F^*_n\,G_n\in \mathbb{C}\,,
\]
but in general
\[
    \langle F,G\rangle =  \lim_{N\to\infty} \langle F, G^N\rangle =  \lim_{N\to\infty}\sum_{n=-N}^N F^*_n\,G_n = \infty\,.
\]
With this in mind in our search for a rigorous description of a quantum mechanics with generalised functions we note the form of the objects
\[
     \langle F_\theta , G_\phi \rangle = \sum_{n=-\infty}^\infty  F^*_n  \, G_n\,e^{in(\phi - \theta )}
\]
and posit the possibility that we can define a generalised sesquilinear product
\begin{equation}\label{defP}
     \langle F_\theta , G_\phi \rangle_{g} := P( F_\theta , G_\phi ) = \sum_{n=-\infty}^\infty  F^*_n  \, G_n\,e^{in(\phi - \theta )}\,e_n\,,\quad \theta,\phi \in [-\pi,\pi ]\,,
\end{equation}
and in particular,
\[
     \langle F , G \rangle_g = \sum_{n=-\infty}^\infty  F^*_n  \, G_n\,e_n\,.
\]
We note that the series in (\ref{defP}) is convergent in the sense of distributions since $F,G\in  \mathcal{D}^\prime (\mathbb{T})$ implies that
\[
     \lim_{\vert n\vert \to\infty} \frac{\vert F_n^*\,e^{-in\theta }\vert}{(1+n^2)^j} = \lim_{\vert n\vert \to\infty} \frac{\vert F_n^*\vert}{(1+n^2)^j} = 0
\]
and
\[
     \lim_{\vert n\vert \to\infty} \frac{\vert G_n\,e^{in\phi\vert}}{(1+n^2)^k} = \lim_{\vert n\vert \to\infty} \frac{\vert G_n\vert}{(1+n^2)^k} = 0
\]
for sufficiently large $j,k\in \mathbb{R}$, hence
\[
     \lim_{\vert n\vert \to\infty} \frac{\vert F_n^* \, G_n\,e^{in(\phi - \theta )}\vert}{(1+n^2)^l} = \lim_{\vert n\vert \to\infty} \frac{\vert F_n^* \, G_n\vert}{(1+n^2)^l} = 0
\]
for some sufficiently large $l\in \mathbb{R}$, because
\[
     \vert F_n^* \, G_n\vert \leqslant \vert F_n^*\vert \, \vert G_n\vert \, .
\]
Hence we can interpret the generalised sesquilinear product $\langle F_\theta , G_\phi \rangle_g$ as an element of $D^\prime (\mathbb{T})$ and note that $P( F_\theta , G_\phi ) = \langle F_\theta , G_\phi \rangle_g \in \mathcal{D}^\prime (\mathbb{T})$ defines a sesquilinear mapping
\[
      P:\,\mathcal{D}^\prime (\mathbb{T})\times\mathcal{D}^\prime (\mathbb{T}) \rightarrow \mathcal{D}^\prime (\mathbb{T})\,.
\]
This mapping is consistent with the $B$-representation in Melsheimer's transformation theory of nonrelativistic quantum mechanics \cite{Melsheimer2}.

As an example we consider the periodic delta-functions $\delta_\theta$ and $\delta_{\theta^\prime}$ that are the angle eigenfunctions of the planar rotator. These have the Fourier expansions
\[
   \delta_\theta (x) = \frac{1}{2\pi} \sum_{n=-\infty}^\infty  e^{-in\theta} e^{inx} \equiv\frac{1}{\sqrt{2\pi}} \sum_{n=-\infty}^\infty  e^{-in\theta} e_n
\]
and
\[
   \delta_{\theta^\prime} (x) = \frac{1}{2\pi} \sum_{m=-\infty}^\infty  e^{-im{\theta^\prime}} e^{imx} \equiv\frac{1}{\sqrt{2\pi}} \sum_{m=-\infty}^\infty  e^{-im{\theta^\prime}} e_m\,,
\]
thus
\[
    \langle \delta_\theta, \delta_{\theta^\prime}\rangle_g = \frac{1}{2\pi} \sum_{n=-\infty}^\infty e^{in\theta}\,e^{-in{\theta^\prime}} e_n = \frac{1}{2\pi} \sum_{n=-\infty}^\infty e^{-in(\theta^\prime -\, \theta)} e_n = \delta_{(\theta^\prime - \, \theta)}(x)\,.
\]
Note that since the delta-functions are even distributions $\delta_\theta (x) = \delta_\theta (-x)$ their generalised sesquilinear form is symmetric:
\[
    \langle \delta_\theta, \delta_{\theta^\prime}\rangle_g = \delta_{(\theta^\prime - \, \theta)}  = \delta_{(\theta - \, \theta^\prime)} = \langle \delta_{\theta^\prime}, \delta_{\theta}\rangle_g\,.
\]
More that this we note that we can write $\delta_\theta (x) = \delta (x-\theta)$ and hence for $x=0$ make the transition from $\theta$ as a label to $\theta$ as a variable that is $\delta(\theta)$. Hence by setting $x=0$ in the generalised sesquilinear product $\langle \delta_\theta, \delta_{\theta^\prime}\rangle_{g} $ we can write the sesquilinear product of delta-functions as
\[
    \langle \delta_\theta, \delta_{\theta^\prime}\rangle_{g} = \delta(\theta^\prime - \, \theta)  = \delta(\theta - \, \theta^\prime) = \langle \delta_{\theta^\prime}, \delta_{\theta}\rangle_{g}\,.
\]


\end{document}